\documentclass[pra,twocolumn,amsmath,amssymb,nofootinbib]{revtex4-2}

\usepackage{graphicx}% Include figure files
\usepackage{dcolumn}% Align table columns on decimal point
\usepackage{bm}% bold math
\usepackage{grfext}% fixes epstopdf options
\usepackage{xcolor}
\usepackage{mathrsfs}
\usepackage[
   colorlinks=true,
   citecolor=blue,
   linkcolor=blue,
   urlcolor=blue
   ]{hyperref}

\newcommand{\ii}{\mathrm{i}}
\newcommand{\ket}[1]{|#1\rangle}

\DeclareMathAlphabet{\boldbBlackboardMath}{U}{BOONDOX-ds}{m}{n}
\newcommand{\Matrix}[4]{% 
  \left(\begin{array}{cc}#1 & #2 \\ #3 & #4\end{array}\right)} 
\newcommand{\Vector}[2]{\left(\begin{array}{c}#1\\#2\end{array}\right)}
\begin{document}

\preprint{Journal}

\title{Quantum Probability and the Born Ensemble}

\author{Themis Matsoukas}
\email{tmatsoukas@icloud.com}
\affiliation{
% \color{brown}{Last Will and Testament II \\ (academically speaking)}
}

\date{\today{\footnotesize ~| Started on December 15, 2022}}
%--------------------------------------------------------------------------*
\begin{abstract}

We formulate a discrete two-state stochastic process with elementary rules that give rise to Born statistics and reproduce the probabilities from the Schrödinger equation under an associated Hamiltonian matrix, which we identify. 
We define the probability to observe a state, classical or quantum, in proportion to the number of \textit{events} at that state---number of ways a walker may materialize at a point of observation at time $t$ through a sequence of transitions starting from known initial state at $t=0$. 
The quantum stochastic process differs from its classical counterpart in that the quantum walker is a pair of qubits, each transmitted independently through all possible paths to a point of observation, and whose recombination may produce a positive or negative event (classical events are never negative). 
We represent the state of the walker via a square matrix of recombination events, show that the indeterminacy of the qubit state amounts to rotations of this matrix, and show that the Born rule counts the number qubits on this matrix that remain invariant over a full rotation. 

\end{abstract}
%--------------------------------------------------------------------------*
\pacs{PACS}
\maketitle
%--------------------------------------------------------------------------*
\section{INTRODUCTION}
Probability has proven remarkably useful in the physical and social sciences but truly indispensable in two major theories in physics: statistical mechanics, and quantum mechanics. Quantum mechanics is special: Unlike statistical mechanics, which wraps probability around the laws of motion, quantum mechanics places probability at the center of these very laws. 
As central as it is to quantum mechanics, probability is not its fundamental mathematical object. This role belongs to the wave function, whose evolution is prescribed by the Schrödinger equation and from which all observables are calculated. The fundamental object of statistical mechanics is not probability either. It is the partition function--\textit{entropy}-- from which all thermodynamic properties are obtained. 
But whereas thermodynamic probability is assigned in proportion to the multiplicity of macrostate, quantum probability is assigned in proportion to the square-amplitude of the wave function, a mathematical prescription with no apparent explanation.
This feature seems to place quantum probability on a level of its own: not merely a property of the state but an intrinsic part of it, which, unlike classical probability, cannot be derived from an ensemble \cite{vanKampen,Wilce:StanfordEncylopediaPhil21}. 
Quantum mechanics is often viewed as a new kind of probability theory that rests on a non-classical propositional logic \cite{Feynman:Proc2ndBerkelySympMathStatProb51,Pitowsky:Book06,Wilce:StanfordEncylopediaPhil21}, one that can be carried outside quantum mechanics, as it has---in quantum computation \cite{Aharonov:arxiv02,Kempe:ContempPhys03,Childs:PhysRevA04}, game theory \cite{Eisert:PRL99,Yukalov:PhilTrans16}, econphysics \cite{Abel:PhilTransRSA23} and elsewhere. 
 
The disconnect with what Vaidman calls ``Nature's equal rule'' \cite{Vaidman:Book20}---the intuitive expectation that probabilities \textit{ought} to arise from \textit{some} equiprobable ensemble---has been a source of tension that has motivated multiple attempts to derive the Born rule from more elementary principles, including quantum logic \cite{Gleason:JMM57,Pitowsky:JMathPhys98}, operational approaches \cite{Caves:FndPhys04,Busch:PhysRevLett03}, symmetry-based arguments \cite{Deutsch:PRSL99,Zurek:PhysRevA05,Wallace:SHPSB07,Sebens:BritishJPhilSci18}, 
stochastic mechanics \cite{Nelson:PhysRev66,Yang:JMathPhys21} and various  reformulations of quantum theory based on alternative postulates \cite{Hardy:arXiv01,Masanes:NatComm19,Auffeves:ArXiv15}. 
A thorough review is given by Vaidman \cite{Vaidman:Book20}. 
Diverse as these approaches may be, they share one feature in common: they \textit{presume} the existence of the wave function within the Hilbert-space structure of quantum mechanics and seek to elucidate its relationship to quantum probability.    
In light of Gleason's theorem \cite{Gleason:JMM57}, which suggests a tight interconnection between all elements of quantum theory \cite{,Pitowsky:JMathPhys98,Fuchs:book11}, the notion that an independent derivation of the Born rule may be obtained from other elements of the theory seems circular. 
Without further assumptions, Vaidman concludes \cite{Vaidman:Book20}, the derivation of the Born rule from existing formulations of quantum mechanics seems impossible. 
Central to the problem is the very notion of quantum probability, whose precise meaning remains shrouded in fog and continues to be debated, not only along the traditional lines that divide frequentist, propensity and Bayesian camps \cite{Jaynes:PhysRev57b,
Jaynes:SFI90,
Ballentine:RevModPhys70,
Appleby:FoundPhys05,
Carroll:QuantaMag19}, but also from the perspective of alternative interpretations and alternative axiomatic formulations of quantum theory \cite{
Rovelli:IntJTheoreticalPhys96,
Deutsch:PRSL99,
Hardy:arXiv01,
Fuchs:RevModPhys13,
Frauchiger:ArXiv17,
Carroll:QuantaMag19}. 
Clearly, no derivation of the Born rule may succeed without a mathematical \textit{definition}---not \textit{interpretation}---of probability. 

Quantum probability, wave function and the Born rule are facets of the same problem and must be tackled simultaneously.  
%,
Since no such puzzle arises in classical probability, the starting point ought to be grounded in lessons learned in the classical domain. 
Our goal is to bridge the gap between classical and quantum probability on the premise that the quantum process in its core is a stochastic process, and that the definition of probability ought to be common to all stochastic processes independently of the particular rules, classical or quantum, that govern them. 
The approach we pursue is fundamentally different from previous attempts. First, we provide a precise definition of probability in the context of the classical stochastic process (CSP).
We do this by formulating the CSP as a walk on the graph that represents all possible  states that can be observed at a given time, 
quantify the multiplicity of observation as the number of paths that connect it to the initial condition, 
and define its probability in direct proportion to this multiplicity. 
This establishes the set of paths that connect an observation at time $t$ to that at $t_0$ as the fundamental ensemble of equiprobable events \cite{Matsoukas:arXivSM23}.  
We then seek suitable modifications to the rules---\textit{not} probabilities--- of the classical stochastic process so as to produce quantum behavior, namely, a quantum stochastic process (QSP) with Born statistics. 

Our development will be within the domain of stochastic processes with just enough input from physics to define what we mean by ``quantum behavior.''  
To this end, we focus on the simplest stochastic process possible, one in which a random variable transitions between two values with time independent rates. Its quantum analogue, frequently employed in other approaches as well \cite{Zurek:PhysicsToday14,Deutsch:PRSL99,Sebens:BritishJPhilSci18}, is the two-state system with time independent Hamiltonian, and is analytically solvable. 
%

%--------------------------------------------------------------------------*
\begin{figure}
  \includegraphics[width=3in]{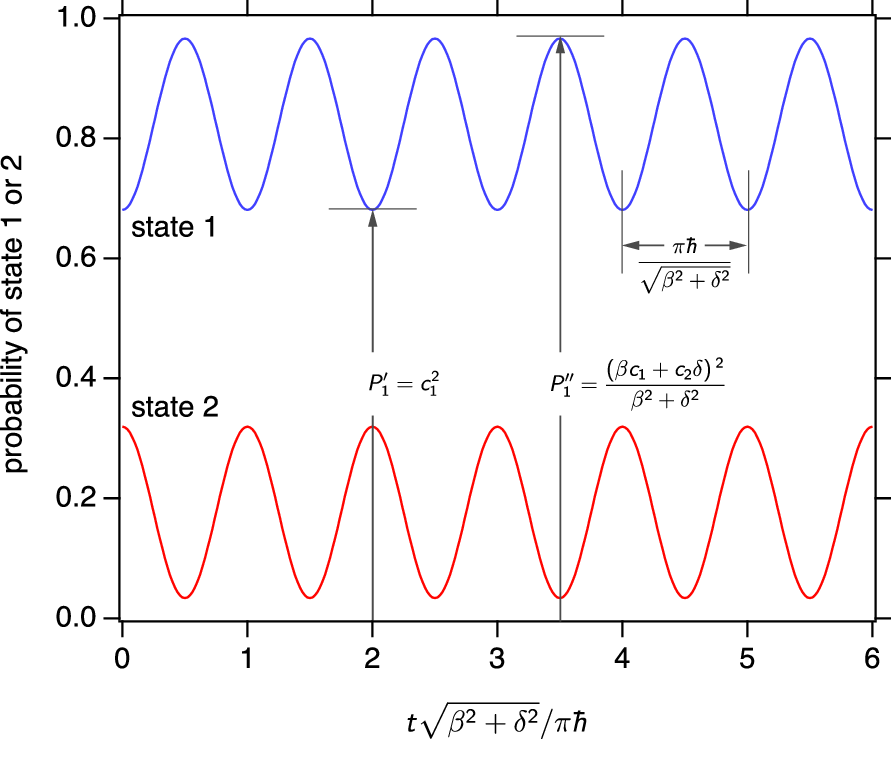}
  \caption{Evolution of probabilities in two-state quantum system with $\alpha=0$, $\beta=1.56$, $\delta=1.255$, $c_1=0.825$.}
\label{fig1}
\end{figure}
%--------------------------------------------------------------------------*

The evolution of the two-state quantum system is governed by the Schrödinger equation,
 \begin{gather}
    \ii\hbar \frac{d \ket{\psi}}{dt} = \mathscr H\, \ket{\psi} , 
 \end{gather}
with real Hamiltonian matrix
\begin{gather}
  \mathscr H = 
  \Matrix
  	{\alpha+\delta}
  	{\beta}
  	{\beta}
  	{\alpha-\delta}
\end{gather}
and initial condition $\ket{\psi(0)} = \binom{c_1}{c_2}$ with $c_1^2+c_2^2 = 1$. 
The probability of state, given by $P_i(t) = \psi_i^*\psi_i^{}$, is oscillatory in time with period $\pi\hbar/\sqrt{\beta^2+\delta^2}$ and amplitude that depends on $\beta$, $\delta$ and the initial state; $\alpha$ makes no contribution to the modulus of the wave function and has no effect on probabilities (see \textcolor{brown}{Appendix \ref{app_2sqs}}).  Figure \ref{fig1} shows the typical behavior. 
 
We treat the probability obtained from the Schrödinger equation as an ``experimental'' result and set out to formulate a discrete stochastic process with rules that quantitatively reproduce its behavior, namely, 
its oscillatory evolution and its dependence on the Hamiltonian matrix and the initial state. This is the only input from quantum theory we will require.  The organization of the paper and the main results are summarized below.
 
In Section \ref{sct:csp} we formulate the classical two-state stochastic process as a walk on the graph of possible transitions, define probability in proportion to the number of paths to reach a state on this graph, and show that the process reaches borderline quantum behavior when the rate of self-transitions is zero. 
 
In Section \ref{sct:qsp} we demonstrate quantum behavior by applying two ad hoc modifications to the classical process: we allow for negative transition rates and use the Born rule to ensure positive probabilities. 
 
In Section \ref{sct:born_rule} we present the main result of the paper: we formulate a \textit{model} for the quantum walk, a set of rules for transmitting the walker along the graph of transitions, from which we \textit{derive} the Born rule. The crucial element is that the quantum walker consists of two qubits that are transmitted independently through the graph of transitions, and whose recombination may produce a positive or negative event. 
 
In Section \ref{sct:born_ensemble} we introduce the event matrix as a representation of the states of the quantum walker, show that the indeterminacy of its state amounts to rotations of this matrix in steps of $\pi/2$, and demonstrate that the Born rule counts the elements of this matrix that are invariant during a full rotation. 
 
In Section \ref{sct:discussion} we summarize the commonalities and differences between the classical and the quantum stochastic process, and discuss the toss of a fair coin as an example that can be treated as either fully classical, or fully quantum stochastic process. 
We offer our concluding remarks in Section \ref{sct:conclusions}.

%--------------------------------------------------------------------------*
\section{Classical two-state stochastic process}
\label{sct:csp}
\begin{figure}
  \includegraphics[width=1.5in]{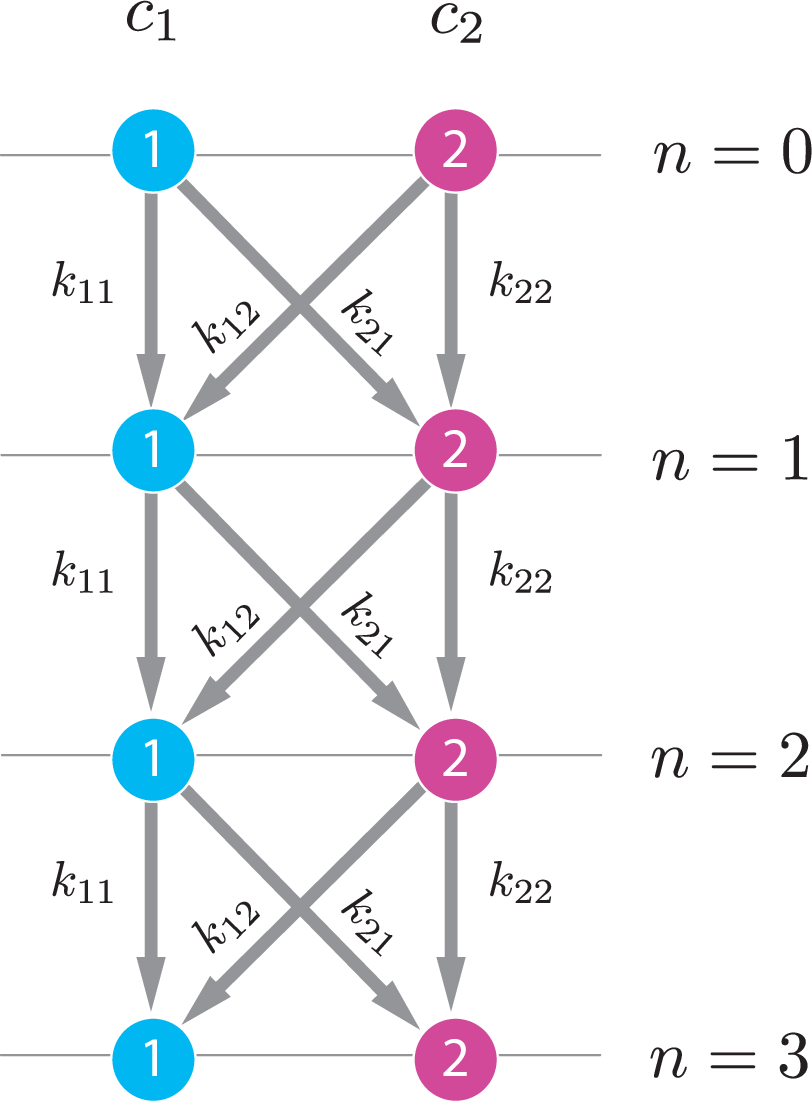}
  \caption{Layered graph of binary transitions. The transition rate $k_{ij}$ is the number of unit paths (channels) from state $j$ to state $i$ in discrete time steps $ \tau$, equal to the mean time between transitions. Node $(i;n)$ represents the observation of state $i=1$ or 2 at $t_n=n\tau$. 
  The multiplicity of observation is the number of paths that arrive at the node starting at the initial state. The probability of observation is proportional to its multiplicity. 
  }
  \label{fig2}
\end{figure}

The classical two-state system is that of two states undergoing transitions of the form $1\leftrightharpoons 2$ with constant transition rates $k_{ij} = k_{i\leftarrow j}$, where $i,j = 1~\text{or}~2$. 
We assume the initial state is known and wish to predict the state following an observation at future time $t_n = n \tau$, $n=1,2\cdots$, where $\tau$ is the mean time between transitions. 
We construct the space of possible observations in steps of $\tau$, and represent it as a graph of states layered by time and directed from past to future (Fig.\ \ref{fig2}). The nodes in the $n$th layer are the possible observation at $t_n$, its edges are transitions that point to all possible observations at $t_{n+1}$. 
By ``observation $(i;n)$'' we refer to the proposition ``the state at time $t_n$ is observed to be $i$, given the last known state at $t=0$,'' where $i$ is 1 or 2  (we return to the notion of ``observation'' in Section \ref{sct:discussion}). 

We view the stochastic process as a walk on this graph and the transition rate $k_{ij}$ as the number of unit channels capable of transmitting the walker from node $i$ to $j$ \cite{Matsoukas:arXivSM23}. Starting from known initial state the walker arrives at observation point $(i;n)$ via a continuous path of interconnected channels. 
The arrival of the walker at a node signifies an event on that node. 
The multiplicity $a_{i;n}$ of observation $(i;n)$ is the number of events consistent with the observation, namely, the number of paths that arrive at that observation in $n$ consecutive transitions from the initial state. 
We define the probability of observation $(i;n)$ as the number of events at the observation normalized by the number of all events at $t_n$:
\begin{gather}
\label{prob_classical}
  P_{i;n} = \frac{a_{i;n}}{a_{1;n} + a_{2;n}},
  \quad i=1,2 .  
\end{gather}
This amounts to assigning equal probability to all events with the same number of transitions. 

The propagation of multiplicity may be viewed as a flow of walkers between nodes: node $(i;n-1)$ with multiplicity $a_{i;n-1}$ delivers $a_{i;n-1} k_{ij}$ walkers to node $(j;n)$; the multiplicity of $(j;n)$ is obtained by summing all flows to that node. The result is expressed concisely in matrix form as
\begin{gather}
\label{propagation}
  \ket{a}_n = \mathscr K \ket{a}_{n-1},
\end{gather}
where $\ket{a}_n= \binom{a_{1;n}}{a_{2;n}}$ is the column vector of multiplicities of all observations at $t_n$, and $\mathscr K$ is the transition matrix
\begin{gather}
  \mathscr K = \Matrix{k_{11}}{k_{12}}{k_{21}}{k_{22}} . 
\end{gather}
By recursive application of Eq.\ (\ref{propagation}) we obtain
\begin{gather}
\label{propagator}
  \ket{a}_n = \mathscr K^n\, \ket{a}_0 .
\end{gather}
Multiplicities are understood to be conditioned on the initial state ($n=0$), which is assumed to be known and whose multiplicity is by convention 1. In the general case the initial condition is $\ket{a}_0=\binom{c_1}{c_2}$ with $c_i\geq 0$.  
Probabilities are calculated by combining Eq.\ (\ref{prob_classical}) with (\ref{propagation}). The stationary probability, if it exists, is obtained by setting $P_{i;n} = P_{i;n-1} = P^*_i$ in Eq.\ (\ref{prob_classical}) and solving for $P^*_i$. 

Equations (\ref{prob_classical})--(\ref{propagator}) summarize the classical stochastic process (CSP) \cite{Matsoukas:arXivSM23} applied to the two-state system. 
The general CSP is not Markov but encompasses the Markov process as a special case. In the Markov process the propagation of probabilities is given by Eq.\ (\ref{propagation}) and its solution by Eq.\ (\ref{propagator}) with all multiplicities replaced by probabilities \cite{Rozanov:77}. 
This replacement is possible if the transition rates satisfy  $k_{11}+k_{21} = k_{12}+k_{22}=1$, in which case the $k_{ij}$ represent properly normalized transition probabilities (see \textcolor{brown}{Appendix \ref{app_markov}}).
Algorithmically, the Markov walker is transmitted from current state at $t_{n-1}$ to an adjacent state at $t_n$ by selecting one unique transition among all available at present state. 
The walker in CSP is transmitted from initial state $t_0$ to a state at $t_n$ by selecting one sequence of $n$ unique transitions among all available at $t_0$. 

%--------------------------------------------------------------------------*
%--------------------------------------------------------------------------*

\begin{figure}
  \includegraphics[width=\linewidth]{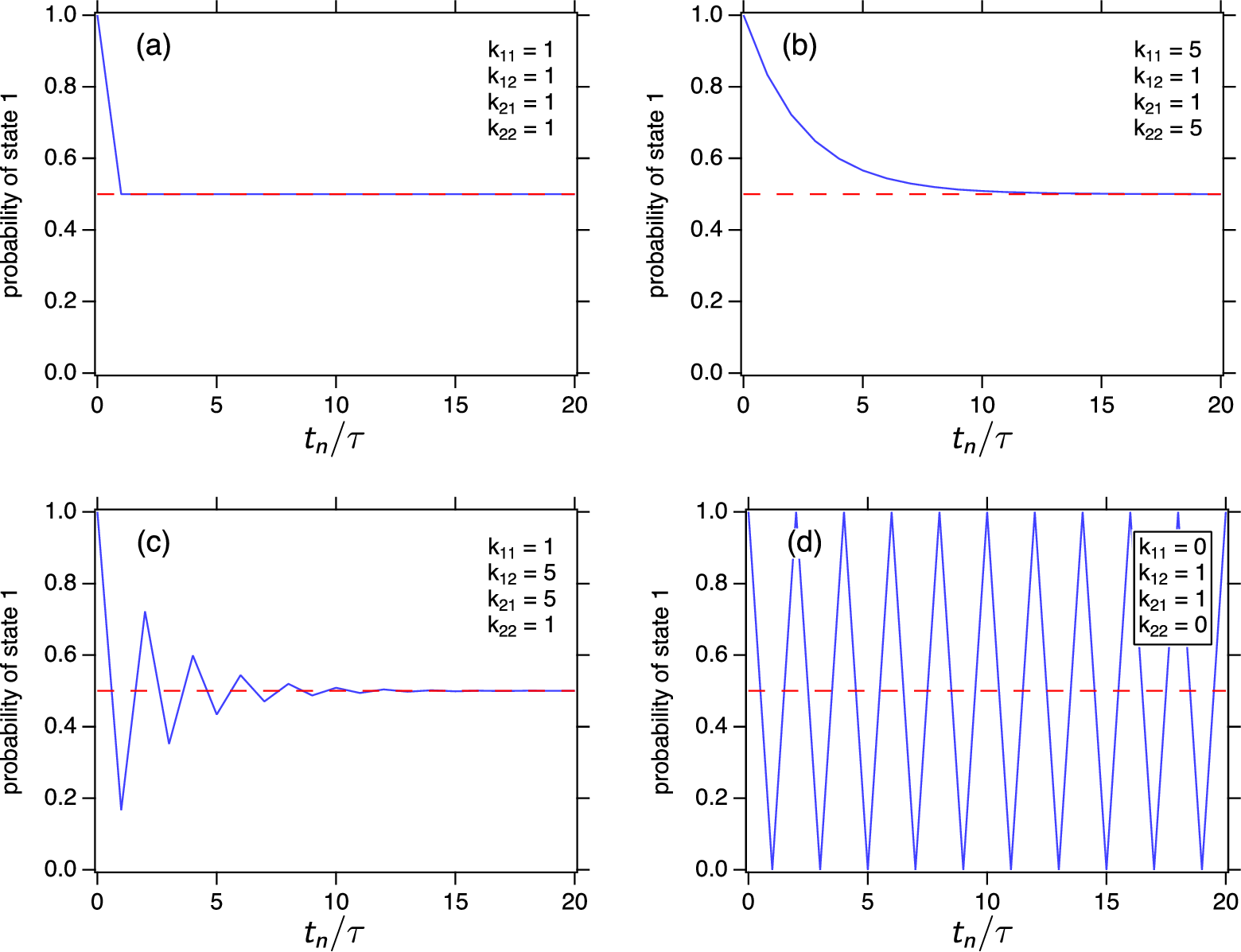} 
\caption{Behaviors of the classical two-state process with initial condition $c_1=1$, $c_2=0$. Except for case (d) the process always reaches a stationary state. }
\label{fig3}
\end{figure}

The two-state CSP exhibits a wide range of dynamics depending on the transition rates. With $k_{11}=k_{21}$, $k_{22}=k_{21}$, the probability reaches its stationary value immediately after the first transition  (Fig.\ \ref{fig3}a). 
If $\text{det}(\mathscr K) >0$ the stationary probability is reached monotonically (Fig.\ \ref{fig3}b); if $\text{det}(\mathscr K) < 0$ it is reached with damped oscillations (Fig.\ \ref{fig3}c).  In the special case $k_{11}=k_{22}=0$, the probability oscillates between $c_1$ and $c_2$ indefinitely (Fig.\ \ref{fig3}c). 
This is borderline-quantum behavior similar to that in Fig.\ \ref{fig1}. However, the amplitude of oscillations is independent of $k_{12}$ or $k_{21}$.

%--------------------------------------------------------------------------*
\section{Quantum stochastic process (QSP)}
\label{sct:qsp}

The classical model reaches borderline quantum behavior when the rate of cross-transitions is reduced to zero, suggesting that negative rates may be the door to the quantum. 
But first we ascribe meaning to negative transition rates:  
If we view the walker as a signal transmitted along the edges of the graph and take that signal to be $+1$, the multiplicity of observation is the total signal arriving at the observation through all possible paths. 
If we suppose that channel $i\to j$ flips the sign of the signal to $-1$, the signal that emerges from the transition is $-|k_{ij}|$; if the signal enters as $-1$, it is transmitted as $|k_{ij}|$. Mathematically this is equivalent to a transition with negative $k_{ij}$. 

We now modify the classical model to allow transition rates to be either positive or negative. The total signal to observation $(i;n)$ is still given by Eq.\ (\ref{propagator}), but as this may now be positive or negative, it no longer represents multiplicity, and probabilities cannot be set proportional to it. To ensure positivity we adopt the quadratic rule
\begin{gather}
\label{prob_Q}
  P_{i;n} = \frac{a^2_{i;n}}{a^2_{1;n}+a^2_{2;n}} .   
\end{gather}
The ad hoc introduction of the Born rule is made at this point for the sole purpose of demonstrating that this model is capable of replicating the behavior of the Schrödinger equation.  The rule itself will be derived from more elementary considerations in Section \ref{sct:born_rule}. 

%--------------------------------------------------------------------------*
%--------------------------------------------------------------------------*

The new model has an even broader range of behaviors than its classical counterpart but the space of interest corresponds to the conditions $k_{11}=-k_{22}$, $k_{22}^2+k_{12} k_{21}\neq 0$. These produce sustained oscillations with amplitude that depends on the transition rates and the initial state (see \textcolor{brown}{Appendix} \ref{app_oscillations}). 
To achieve full equivalence between the  QSP and the two-state quantum system we match the period and amplitude of the oscillations between the two models for all $c_1^2+c_2^2=1$ and obtain the following relationships between the transitions rates of the QSP and the elements of the Hamiltonian (see \textcolor{brown}{Appendix} \ref{app_equivalence}): 
\begin{gather}
% \label{period}
%   \tau =  \pi\hbar/2\sqrt{\beta^2+\delta^2},\\
\label{equiv}
  k_{11} = - k_{22} = \lambda \delta,\quad
  k_{12} =   k_{21} = \lambda \beta , 
\end{gather}
where $\lambda\neq 0$ is arbitrary and may be positive or negative. 
Thus we arrive at the relationship between the transition matrix of the QSP and the equivalent Hamiltonian:
\begin{gather}
  \mathscr K = \lambda\left(
    \mathscr H - \alpha I
  \right),
\end{gather}
where $I$ is the identity matrix. 
With $\lambda=1$,  $\alpha=0$, this amounts to $\mathscr K = \mathscr H$. 
We confirm by calculating the probabilities of the two-state quantum system using the transition matrix of the QSP as its Hamiltonian and plot the solution against the probabilities obtained by QSP in Fig.\ \ref{fig_Q}b. 
The QSP samples the solution of the Schrödinger equation at discrete intervals of half-period. Conversely, the Schrödinger equation produces a continuous interpolating function for the probability of the discrete QSP. This interpolating function may in fact be derived independently of the Schrödinger equation (see \textcolor{brown}{Appendix} \ref{app_wavefunction}). 
If we adopt the normalization $k_{11}^2+k_{12}^2 = k_{21}^2 + k_{22}^2 = 1$, the transition matrix becomes unitary and probabilities reduce to $P_{i;n} = a_{i;n}^2$ (see \textcolor{brown}{Appendix} \ref{app_unitary}).

%--------------------------------------------------------------------------*
\section{Born Rule}
\label{sct:born_rule}

We now seek to formulate a model that \textit{gives rise} to the Born rule, i.e., prescribe a set of rules for the quantum walker such that the quantity $a_{i;n}^2$ counts the number of equiprobable events at observation $(i;n)$.  
The hint comes from the Born rule itself, whose quadratic dependence on the transmitted signal suggests a binary event per walker, as if twice as many signals as walkers are transmitted to an observation, where they recombine in all possible pairs. 
Based on this idea we formulate the following model:

%--------------------------------------------------------------------------*
\begin{figure}
  \includegraphics[width=\linewidth]{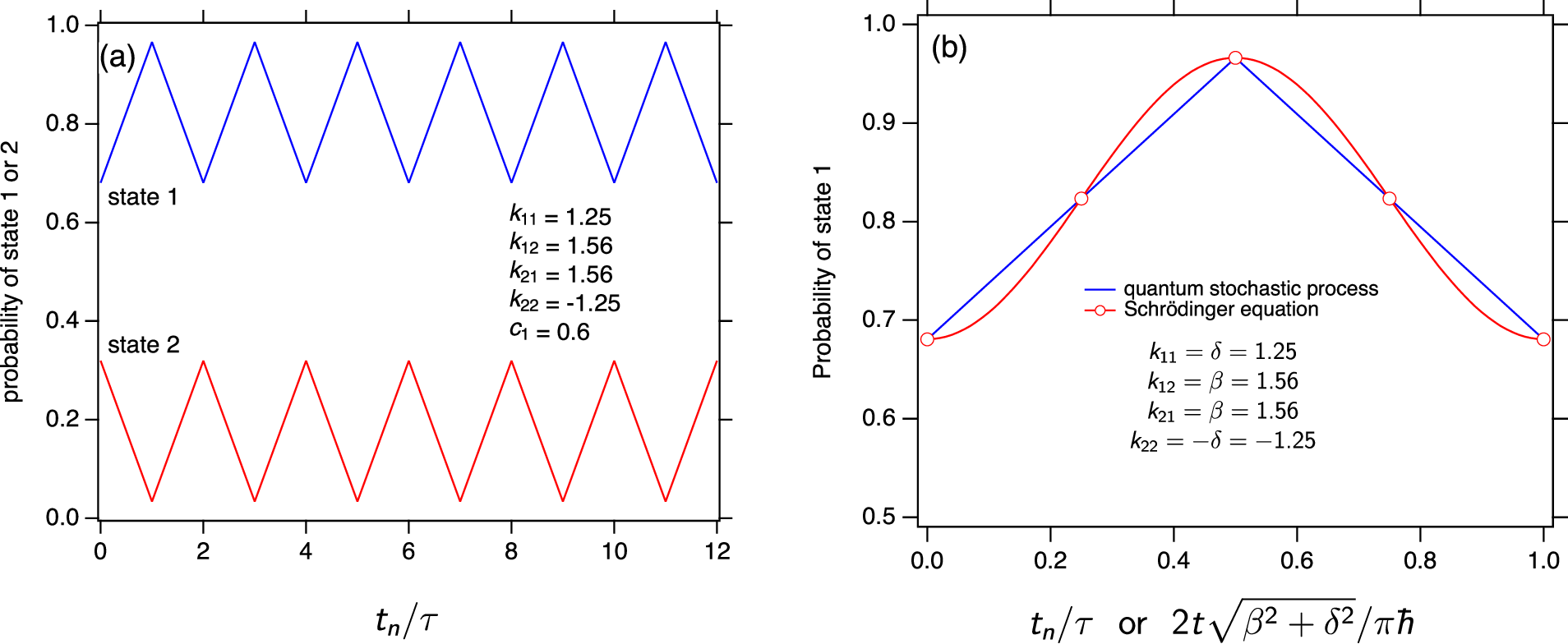} 
  \caption{(a) Two-state stochastic process with $k_{22}=-k_{11}$, $k_12=k_{21}$,  produces sustained oscillations. (b) The quantum stochastic process samples the Schrödinger solution with $\alpha=0$, $\mathscr H=\mathscr K$. }
\label{fig_Q}
\end{figure}
%--------------------------------------------------------------------------*

\begin{enumerate}
  \item The quantum walker is a pair of elements---qubits---that exist in one of two  equivalent states. We represent them as ``type up'' ($\uparrow$) and ``type down'' ($\downarrow$). In the absence of a preferential direction, to use the arrow analogy, the state of a single qubit cannot be declared to be up or down, it is merely an arrow. On the other hand, two qubits may be compared and the pair may be found to be either in-phase, if both qubits point in the same direction, or out-of-phase, if they point in opposite directions. 

  \item The qubits that make up the walker are transmitted independently of each other, each following its own path on the graph of transitions. When a qubit passes through a positive channel it is transmitted as is; when it passes through a negative channel its type is flipped.

  \item When qubits recombine at a point of observation they reconstitute the walker either in its initial state (in-phase or out-phase), in which case the recombination counts as a positive event, or in its opposite state (in-phase walker is reconstituted as out-of-phase, and vise versa), in which case the event is negative. 

  \item The multiplicity of observation is the net number of events that materialize at the point of observation through all possible paths, and is equal to the difference between the number of positive and negative events at the observation. This difference will be shown to never be negative. 

  \item The probability of observation is its multiplicity normalized by the multiplicity of all observations at the same time. 

\end{enumerate}

%--------------------------------------------------------------------------*
\begin{figure}[t]
  \includegraphics[width=\linewidth]{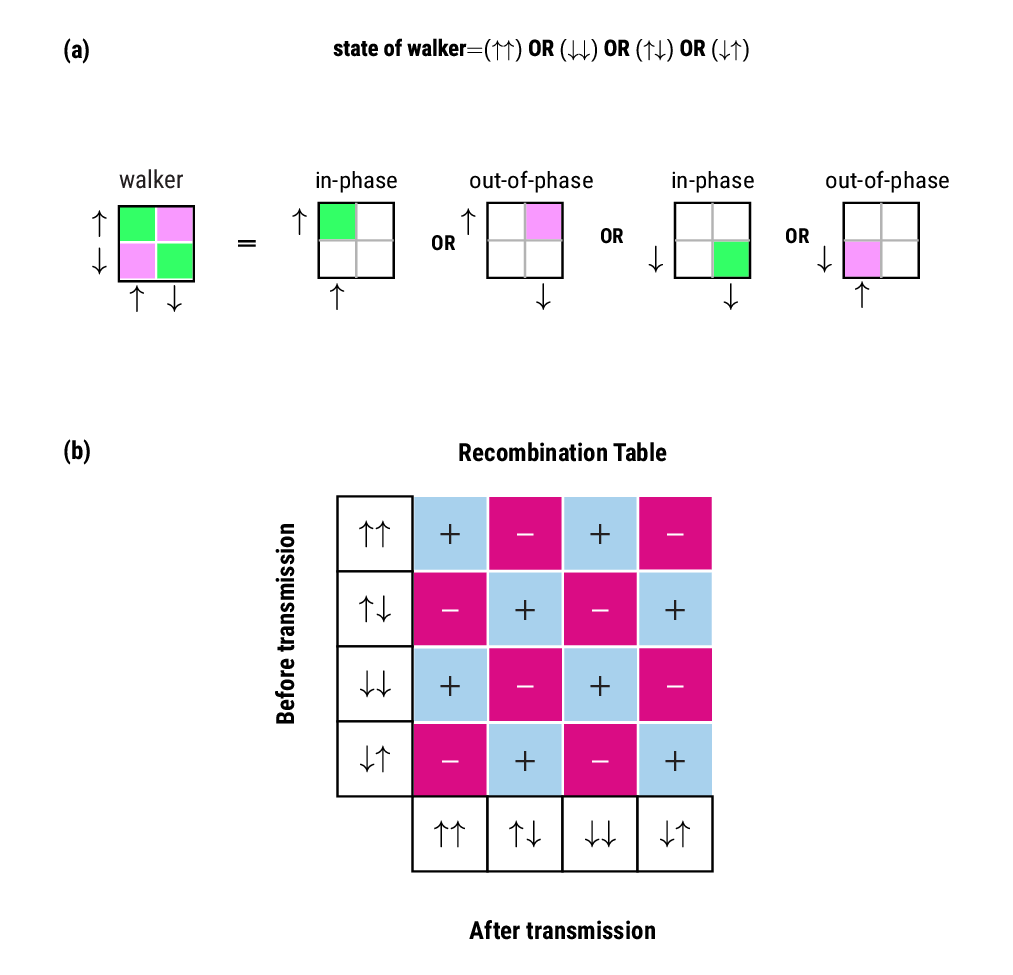}  
  \caption{(a) The state of the quantum walker is a pair of qubits that are either in-phase  or out-of-phase. In matrix form these states can be arranged as a sequence of rotations by $\pi/2$.  (b) When the walker is reconstituted after transmission the event is positive if states are the same before and after transmission (both in-phase or both out-of-phase), and negative otherwise. }
\label{fig5}
\end{figure}
%--------------------------------------------------------------------------*

%--------------------------------------------------------------------------*
\begin{figure}[t]
  \includegraphics[width=\linewidth]{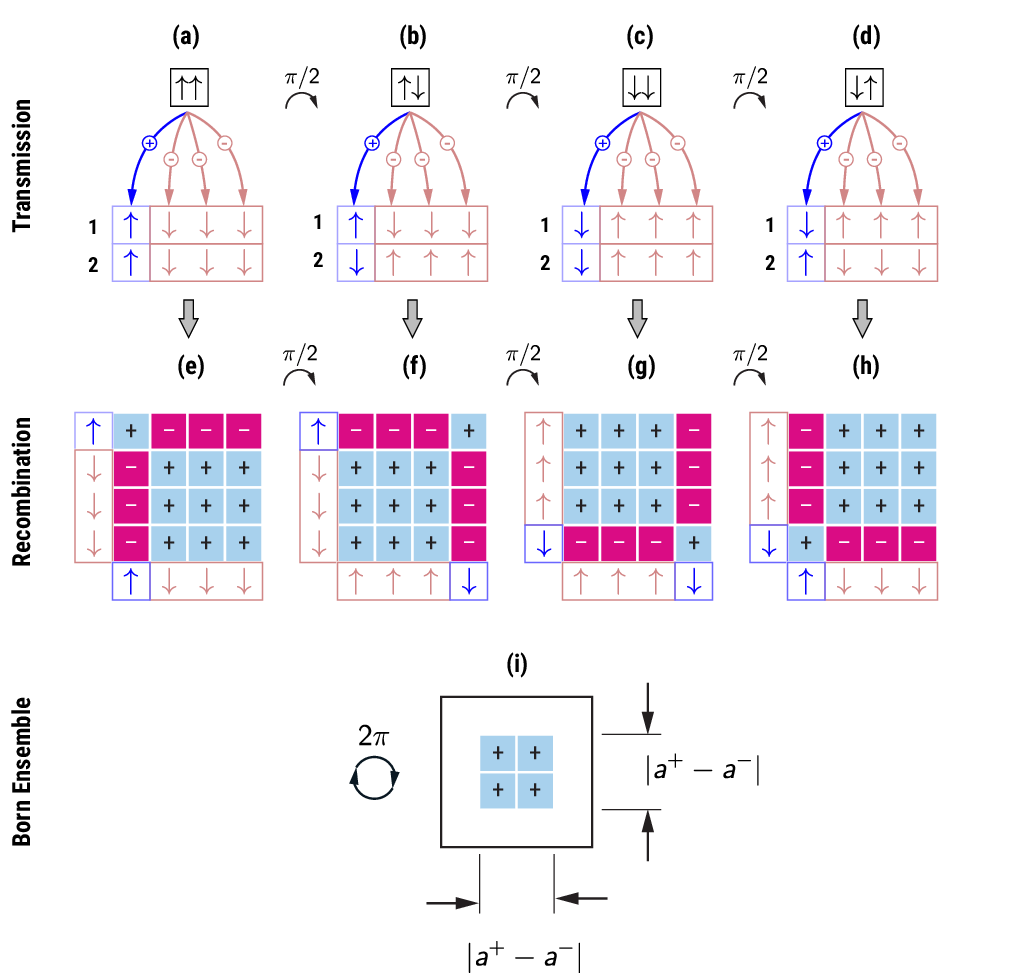} 
  \caption{
  (a--d) Transmission of quantum  walker through $a^+$ positive (\textcolor{blue}{blue}) and $a^-$ negative (\textcolor{brown}{brown}) paths (1 and 2 refer to the state of the first and second qubit of the walker after transmission). Positive paths transmit the qubit as is, negative paths invert it. 
  (e--h) The event matrix is constructed by pairing the transmitted qubits in all possible ways. The elements of the matrix have a value of $+1$ if both elements of the walker have been transmitted through paths with the same sign, and $-1$ otherwise. Each rotation of the walker before transmission corresponds to a rotation of the event matrix after transmission. 
  (i) The mean event matrix over four rotations forms the Born ensemble, a symmetric matrix with exactly $\mathscr B=(a^+-a^-)^2$ unit elements. These are the elements that remain invariant during rotation.  In this example $a^+=1$, $a^-=3$, $\mathscr B = 4$. } 
\label{fig6}
\end{figure}
%--------------------------------------------------------------------------*
These rules describe a quantum stochastic process in which walkers may annihilate each other at the point of observation. 
As in the classical case, the probability of observation is proportional to the total number of events at the observation. 
The calculation of this number is straightforward. Let $a^+_{i;n}$ be the number of paths arriving at observation $(i;n)$ with an even number of flips (positive paths) and $a^-_{i;n}$ the number of paths with odd number of flips (negative paths). Their difference,
\begin{gather}
 a^+_{i;n} - a^-_{i;n} =  a_{i;n} , 
\end{gather}
is the total signal arriving at $(i;n)$ and is given by Eq.\  (\ref{propagation}); it may be positive or negative. 
If two qubits both travel through positive paths, they reconstitute the walker in its original state and the event is positive. The same is true if they both travel through negative paths. The number of positive recombinations is $(a^+_{i;n})^2 + (a^-_{i;n})^2$. 
If the two qubits travel through paths of opposite sign, recombination flips the state of the pair and the event is negative; the number of negative combinations is $2 a^+_{i;n} a^-_{i;n}$. The difference between positive and negative recombinations is
\begin{multline}
\label{born_rule}
  (a^+_{i;n})^2 + (a^-_{i;n})^2 - 2 a^+_{i;n} a^-_{i;n} 
  \\ 
  = (a^+_{i;n} - a^-_{i;n})^2 = a_{i;n}^2
  \equiv \mathscr B_{i;n}  , 
\end{multline}
and is never negative.  

\begin{quote}\em
  This is the Born rule of quantum probability; we have obtained it by straightforward counting of the number of events that materialize at an observation.  
\end{quote}
We refer to the net number of events as the Born number $\mathscr B_{i;n}$ of the observation. 
We should make a note of the fact that the symmetry of transmission in the forward and reverse direction of the graph implies that we might as well take one or both qubits to travel from the observation to the initial state. We will continue, however, to view the QSP walker as two qubits both transmitted forward in time. 

To provide context to these rules we begin by noting that the quantum walker is assembled by pairing two qubits in all combinations of states. This pairing is shown in Fig.\ \ref{fig5}a as a 2$\times$2 matrix with the in-phase pairs ($\uparrow\uparrow$, $\downarrow\downarrow$)  on one diagonal and the out-of-phase pairs ($\uparrow\downarrow$, $\downarrow\uparrow$) on the other.  
The four configurations of the pair can be regarded as a sequence of matrix rotations in steps of $\pi/2$. 
We adopt the view that the state of the walker, just as that of its qubits, cannot be determined in isolation and may only be assessed in comparison with another walker. 
The states of two walkers are either the same (in-phase/in-phase or out-of-phase/out-of-phase) or different (in-phase/out-of-phase or out-of-phase/in-phase). Each option has a multiplicity of 8, the number of ways to assign a state to individual qubits. Externally, however, we may only detect the binary outcomes ``same'' or ``different'' (Fig.\ \ref{fig5}b). 
In this respect, the association of positive or negative events with walkers that emerge from transitions as ``same'' or ``different'' has its origin in the indeterminacy of the state of the walker. 

The walk described here is quite different from the similarly named discrete-time quantum walk (DTQW) in quantum computation \cite{Childs:PhysRevA04}. 
The walker in DTQW is a single qubit that transitions between adjacent sites under the combined action of a unitary operation that rotates the qubit, and a shift operator that moves it in state space with probabilities according to the Born rule. 
This walk is designed to disperse its walkers in state space faster than the classical walk. 
The QSP involves no such operators, nor does it force the Born rule. Instead, it transmits a pair of qubits over paths composed of any number of transitions, with probability determined by the rule of recombination events.

%--------------------------------------------------------------------------*
\section{Born Ensemble}
\label{sct:born_ensemble}
We have seen that the state of the walker before transmission is represented as a sequence of four matrix rotations in steps of $\pi/2$.
We now show that the state of the walker after recombination is also represented by four rotations, this time of a matrix whose elements are the recombination events, positive and negative, at the observation (event matrix).  
 
Whether the walker emerges from a transition as ``same'' or ``different'' depends on the number of positive and negative transmissions between the end states. 
At the point of recombination we may have either $a^+$ qubits of type $\uparrow$ and $a^-$ qubits of type $\downarrow$, or $a^-$ qubits of type $\uparrow$ and $a^+$ qubits of type $\downarrow$, depending on the rotation of the walker before transmission (Fig.\ \ref{fig6}a--d). The walker is then reconstituted by pairing these qubits in all combinations. 
The pairing is shown as a matrix in Fig.\ \ref{fig6}e-h: on each axis we place  all qubits of type $\uparrow$ followed by those of type $\downarrow$ (the order does not matter but should be kept fixed on each axis throughout this construction) and assign either $a^+$ or $a^-$ qubits to one type, and either $a^-$ or $a^+$ to the other.  The assignments on each axis of the matrix are made  independently of each other. 
The elements of the matrix are assigned a value according to the recombination rule: $+1$ if both qubits travel through paths of the same sign, and $-1$ otherwise.  The result is the \textit{event matrix}: a square matrix with $(a^++a^-)^2$ elements whose sum is $(a^+-a^-)^2$. 
The four ways to assign states to $a^+$ and $a^-$ qubits on each axis produce four rotations of the event matrix in steps of $\pi/2$, each representing the transmission of a rotated state of the walker before transmission (Fig.\ \ref{fig6}e--h). 

There are two types of cells on the event matrix: those that have a value of $+1$ in all four rotations, and those whose value is $+1$ in two rotations and $-1$ in the other two. 
The average event matrix over a full rotation reveals a simple pattern: a square submatrix at the center of the event matrix with $\mathscr B=(a_+-a_-)^2$ unit elements surrounded by zeros (Fig.\ \ref{fig6}i). 
This is the \textit{Born ensemble}: a set of equiprobable events compatible with the observation. 
Events outside this ensemble cancel out because of equal numbers of positive and negative events. 
They represent fluctuations, unstable events that flicker but do not materialize, due to the indeterminate state of the quantum walker.  
Only events inside the Born ensemble---the ``bull's eye'' of the event matrix---contribute to the probability of observation. These events do not fluctuate but remain undisturbed by rotation. 
In this sense the Born rule counts the number of events whose identity remains unaffected by the indeterminate state of the quantum walker.   

%--------------------------------------------------------------------------*
\section{Discussion}
\label{sct:discussion}

To place classical and quantum stochastic processes under a common language we re-interpret the stochastic process as the transmission of a \textit{signal} along a graph that represents the possible transitions of a random variable as a function of time. The transmitted signal propagates according to Eq.\ (\ref{propagation}) and satisfies Eq.\ (\ref{propagator}). 
Multiplicity, defined as the number of events compatible with an observation, is a function of the transmitted signal and cannot be negative. 
What constitutes an ``event'' is specified by the model that governs the walk. 
The operational difference between classical and quantum stochastic process lies in the functional relationship between multiplicity and signal:  classical multiplicity is equal to the transmitted signal itself; quantum multiplicity is equal to the transmitted signal-squared. 
Probability, classical or quantum, is normalized multiplicity---``Nature's equal rule'' restored.  
The wave function of quantum mechanics is the normalized signal $a_{i;n}$ transmitted to the observation. Its classical counterpart is positive or zero--but never negative. The total classical signal at $t_n$ (normalizing factor of probability) is the partition function of the stochastic process \cite{Matsoukas:arXivSM23}; thus wave function and partition function are more closely related than one might have thought. 

Quantum probability is no different from classical probability in any fundamental sense. 
Both are defined in proportion to the number of events consistent with the observation, and in both cases this number calculated from a model that codifies our knowledge about the system. 
What makes quantum probability non intuitive is the possibility of annihilation between positive and negative events. This behavior has no classical counterpart. 
A classical walker that arrives at a point of observation at time $t_n$ may be thought to have arrived from some other state at $t_{n-1}$. 
Any unique sequence of states visited by the walker can be interpreted as an equiprobable trajectory, an evolution of states in time, and observation may be viewed to reveal a pre-existing, albeit stochastic, state on such sequence. 
The same cannot be said for the quantum process. Negative events do not represent real states, and a sequence of recombinations, which may include negative events, cannot represent a sequence of pre-existing states. 
To gain better insight as to why this is so we consider a simple model, a classical process that borders the quantum---and a quintessential example of stochastic behavior: the tossing of a fair coin. 

%--------------------------------------------------------------------------*
\begin{figure}[t]
  \includegraphics[width=\linewidth]{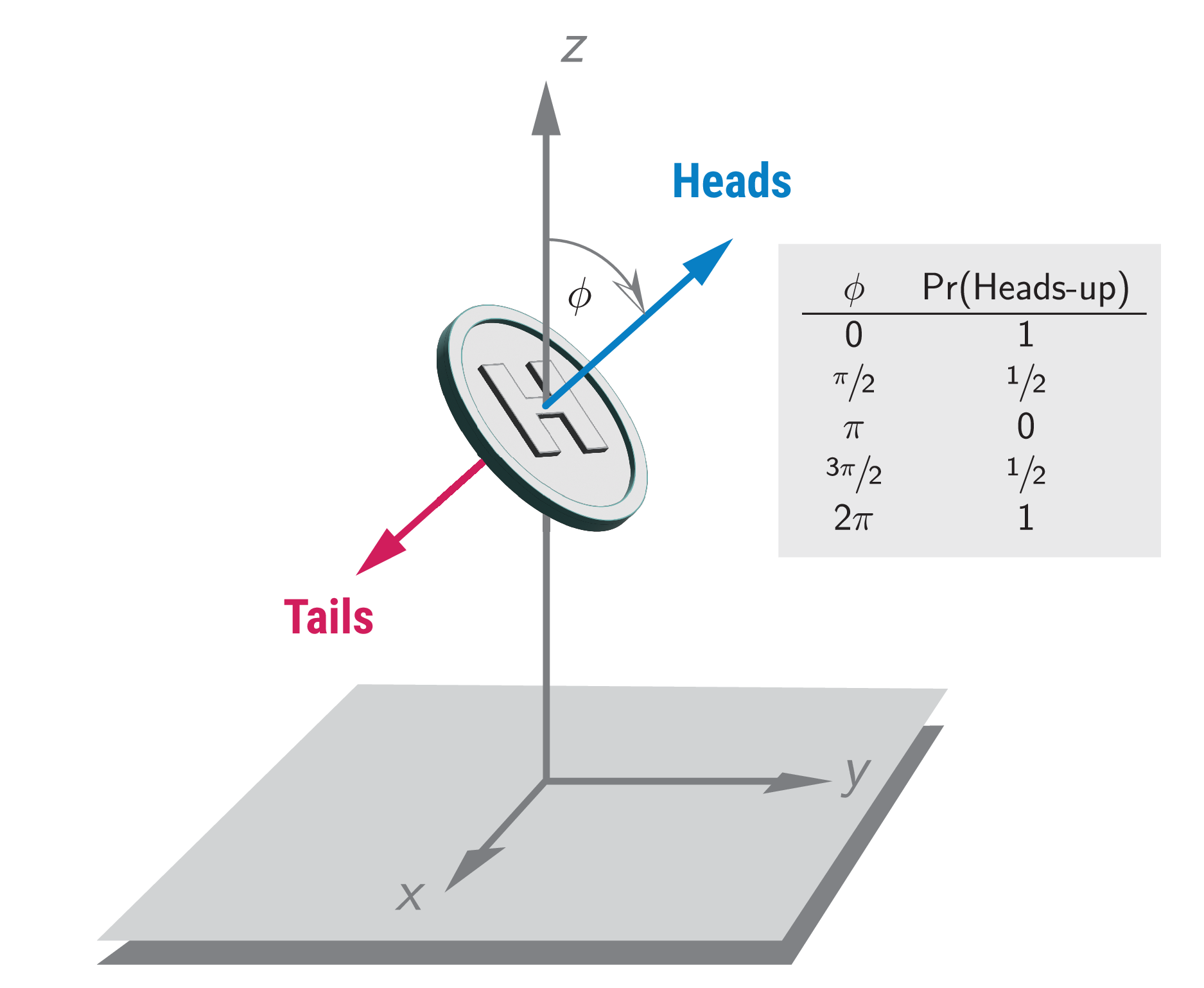} 
  \caption{A fair coin impacts the $xy$ plane at an angle $\phi$ between $z$ and the vector normal to the surface Heads. Probabilities at impact angles $\phi=0$, $\pi/2$, $\pi$ and $3\pi/2$ are assigned based on symmetry arguments: if the coin impacts in perfect alignment with the plane ($\phi=0$ or $\pi$), it will most likely come to rest as is; if it hits the plane exactly on its edge ($\phi=\pi/2$ or $3\pi/2$), it has an equal probability of coming to  rest as Heads-up or Tails-up. These probabilities can be represented either by a classical transition matrix with no negative elements ($k_{11}=k_{22}=0$, $k_{12}=k_{21}\neq 0$), or by a quantum transition matrix  ($k_{11}=-k_{22}\neq 0$, $k_{12}=k_{21}=0$).  
   } 
\label{fig7}
\end{figure}
%--------------------------------------------------------------------------*

We construct a very rudimentary model of a fair coin landing on a surface (Fig.\ \ref{fig7}). We assume that the probability that the coin lands Heads-up or Tails-up depends only on the angle $\phi$ at which the coin hits the $xy$ plane, as measured from the $z$ axis to the vector normal to the side Heads. 
At impact angle $\phi=2\pi n$ the coin hits the table Heads-up and we assume it comes to rest in this position with probability 1; at $\phi=(2n+1)\pi$ it hits the table Tails-up and we take the probability of Heads-up to be 0; at $\phi=\pi/2+n \pi$ the coin hits the table exactly on its edge and we assume that it comes to rest as Heads-up or Tails-up with equal probability. 
These probabilities correspond to the classical transition matrix $k_{11}=k_{22} = 0$, $k_{12}=k_{21} = 1$ with $\tau=\pi$, $c_1=1$, $c_2=0$, shown in Fig.\ \ref{fig3}d. 
But these are precisely the same probabilities obtained from the Schrödinger equation with $\ket{1}=\ket{\text{Heads}}$, $\ket{2}=\ket{\text{Tails}}$, $\delta=0$, $\phi=2\pi\hbar/\beta$, $c_1=1$, $c_2=0$ (see \textcolor{brown}{Appendix \ref{app_wavefunction}}): 
\begin{gather}
\label{fair_coin}
   \Pr(\text{Heads-up}) = \cos^2\frac{\phi}{2},\,
   \Pr(\text{Tails-up}) = \sin^2\frac{\phi}{2}.  
\end{gather}
This model literally sits at the border between classical and quantum: it can be described as either fully classical (non negative $k_{ij}$, probabilities proportional to transmitted signal) or fully quantum ($k_{11}=-k_{22}$, probabilities proportional to signal-square).
 
These probabilities characterize the state of the coin following an ``observation,'' by which we mean ``the side of the coin that points in the $z$ direction when the coin has come to rest after hitting the $xy$ plane.'' 
They do not characterize its instantaneous state, which refers to the orientation of the free coin in space. Observation requires impact with the surface and this, by definition, places the intrinsic state beyond observation (except for $\phi=n\pi$). 
It is for this reason that we have insisted throughout this development on speaking of the probability of \textit{observation} rather than the probability of \textit{state}. The two are not always the same. 
Once an observation is made the coin is found to be either Heads-up or Tails-up. This amounts to a projection of the instantaneous state onto the observation plane that forces the coin, through the physical forces of impact, to one of the two possible states compatible with that plane.  

%--------------------------------------------------------------------------*
\section{Conclusions}
\label{sct:conclusions}

We have constructed a discrete stochastic process with rules that reproduce the wave function and the probability obtained from the Schrödinger equation and the Born rule. 
This process stands on its own: it uses its own primitive concepts and makes no reference to the axioms of quantum mechanics or its mathematical structure.
The ontology it suggests remains an open question but the model offers tangible elements---qubits, recombination, negative events---to be explored. 
We can now construct simple but rigorous quantum stochastic models as instruments of study in pursuit of this exploration. 
By building them in the discrete domain we gain the advantage of bypassing the mathematical distractions of Hilbert spaces, Hermitian operators and complex algebra, so that we may focus on those concepts that might explain quantum behavior. 
Lastly, by recognizing stochastic processes to posses a core structure of such generality that it encompasses the domains of classical and quantum, we may leverage our classical intuition to develop and grow our understanding of the logic that governs the quantum world. 

%--------------------------------------------------------------------------*
\begin{acknowledgments}
  I wish to thank Cal Abel for stimulating discussions on the application of quantum mechanics in econophysics \cite{Abel:PhilTransRSA23} that piqued my interest in quantum probability and its relation to stochastic processes; and Milton Cole for reviewing a draft of this paper.  
\end{acknowledgments}

%--------------------------------------------------------------------------*
% \clearpage
%  apsrev4-2, ieeetr, plain, alpha, acm, siam, elsarticle-num
%  hieeetr   <-- compact, hacked for arxiv
%  apsrev4-2 <-- recommended after cleaning up bib entries
% \bibliography{statMech, tm}

\begin{thebibliography}{10}

\bibitem{vanKampen}
N.~G. van Kampen, {\em Stochastic Processes in Physics and Chemistry}.
\newblock Amsterdam: North Holland Personal Librry, 1990.

\bibitem{Wilce:StanfordEncylopediaPhil21}
A.~Wilce, ``{Quantum Logic and Probability Theory},'' in {\em The {Stanford}
  Encyclopedia of Philosophy} (E.~N. Zalta, ed.), Metaphysics Research Lab,
  Stanford University, {F}all 2021~ed., 2021.

\bibitem{Feynman:Proc2ndBerkelySympMathStatProb51}
R.~P. Feynman, ``The concept of probability in quantum mechanics,'' in {\em
  Proceedings of the Second Berkeley Symposium on Mathematical Statistics and
  Probability: Held at the Statistical Laboratory, Department of Mathematics,
  University of California, July 31-August 12, 1950} (J.~Neyman, ed.),
  University of California Press, 1951.

\bibitem{Pitowsky:Book06}
I.~Pitowsky, ``Quantum mechanics as a theory of probability,'' in {\em The
  Western Ontario Series in Philosophy of Science}, pp.~213--240, Springer
  Netherlands, 2006.

\bibitem{Aharonov:arxiv02}
D.~Aharonov, A.~Ambainis, J.~Kempe, and U.~Vazirani, ``Quantum walks on
  graphs,'' 2002, quant-ph/0012090.

\bibitem{Kempe:ContempPhys03}
J.~Kempe, ``Quantum random walks: An introductory overview,'' {\em Contemporary
  Physics}, vol.~44, no.~4, pp.~307--327, 2003.

\bibitem{Childs:PhysRevA04}
A.~M. Childs and J.~Goldstone, ``Spatial search by quantum walk,'' {\em Phys.
  Rev. A}, vol.~70, p.~022314, Aug 2004.

\bibitem{Eisert:PRL99}
J.~Eisert, M.~Wilkens, and M.~Lewenstein, ``Quantum games and quantum
  strategies,'' {\em Phys. Rev. Lett.}, vol.~83, pp.~3077--3080, Oct 1999.

\bibitem{Yukalov:PhilTrans16}
V.~I. Yukalov and D.~Sornette, ``Quantum probability and quantum
  decision-making,'' {\em Philosophical Transactions of the Royal Society A:
  Mathematical, Physical and Engineering Sciences}, vol.~374, no.~2058,
  p.~20150100, 2016.

\bibitem{Abel:PhilTransRSA23}
C.~R. Abel, ``The quantum foundations of utility and value,'' {\em
  Philosophical Transactions of the Royal Society A: Mathematical, Physical and
  Engineering Sciences}, vol.~381, no.~2252, p.~20220286, 2023.

\bibitem{Vaidman:Book20}
L.~Vaidman, {\em {Quantum, Probability, Logic: The Work and Influence of Itamar
  Pitowsky}}, ch.~26. Derivations of the Born Rule, pp.~567--584.
\newblock Studies in Philosophy and History of Science, Israel: Springer
  Nature, 2020.

\bibitem{Gleason:JMM57}
A.~M. Gleason, ``Measures on the closed subspaces of a {H}ilbert space,'' {\em
  Journal of Mathematics and Mechanics}, vol.~6, no.~6, pp.~885--893, 1957.

\bibitem{Pitowsky:JMathPhys98}
I.~Pitowsky, ``{Infinite and finite Gleason's theorems and the logic of
  indeterminacy},'' {\em Journal of Mathematical Physics}, vol.~39,
  pp.~218--228, 01 1998.

\bibitem{Caves:FndPhys04}
C.~M. Caves, C.~A. Fuchs, K.~K. Manne, and J.~M. Renes, ``{G}leason-type
  derivations of the quantum probability rule for generalized measurements,''
  {\em Foundations of Physics}, vol.~34, no.~2, pp.~193--209, 2004.

\bibitem{Busch:PhysRevLett03}
P.~Busch, ``Quantum states and generalized observables: A simple proof of
  {G}leason's theorem,'' {\em Phys. Rev. Lett.}, vol.~91, p.~120403, Sep 2003.

\bibitem{Deutsch:PRSL99}
D.~Deutsch, ``Quantum theory of probability and decisions,'' {\em Proceedings
  of the Royal Society of London. Series A: Mathematical, Physical and
  Engineering Sciences}, vol.~455, no.~1988, pp.~3129--3137, 1999.

\bibitem{Zurek:PhysRevA05}
W.~H. Zurek, ``Probabilities from entanglement, {B}orn's rule
  ${p}_{k}={\ensuremath{\mid}{\ensuremath{\psi}}_{k}\ensuremath{\mid}}^{2}$
  from envariance,'' {\em Phys. Rev. A}, vol.~71, p.~052105, May 2005.

\bibitem{Wallace:SHPSB07}
D.~Wallace, ``{Quantum probability from subjective likelihood: Improving on
  Deutsch's proof of the probability rule},'' {\em Studies in History and
  Philosophy of Science Part B: Studies in History and Philosophy of Modern
  Physics}, vol.~38, no.~2, pp.~311--332, 2007.

\bibitem{Sebens:BritishJPhilSci18}
C.~T. Sebens and S.~M. Carroll, ``Self-locating uncertainty and the origin of
  probability in {E}verettian quantum mechanics,'' {\em The British Journal for
  the Philosophy of Science}, vol.~69, no.~1, pp.~25--74, 2018.

\bibitem{Nelson:PhysRev66}
E.~Nelson, ``Derivation of the {S}chr\"odinger equation from {N}ewtonian
  mechanics,'' {\em Phys. Rev.}, vol.~150, pp.~1079--1085, Oct 1966.

\bibitem{Yang:JMathPhys21}
J.~M. Yang, ``{Variational principle for stochastic mechanics based on
  information measures},'' {\em Journal of Mathematical Physics}, vol.~62,
  p.~102104, 10 2021.

\bibitem{Hardy:arXiv01}
L.~Hardy, ``Quantum theory from five reasonable axioms,'' 2001,
  quant-ph/0101012.

\bibitem{Masanes:NatComm19}
L.~Masanes, T.~D. Galley, and M.~P. M{\"u}ller, ``The measurement postulates of
  quantum mechanics are operationally redundant,'' {\em Nature Communications},
  vol.~10, no.~1, p.~1361, 2019.

\bibitem{Auffeves:ArXiv15}
A.~Auff{\`e}ves and P.~Grangier, ``A simple derivation of {B}orn's rule with
  and without {G}leason's theorem,'' 2015, 1505.01369.

\bibitem{Fuchs:book11}
C.~A. Fuchs, {\em {Letters to Gilles Brassard}}, pp.~90--104.
\newblock Cambridge University Press, 2011.

\bibitem{Jaynes:PhysRev57b}
E.~T. Jaynes, ``Information theory and statistical mechanics. ii,'' {\em Phys.
  Rev.}, vol.~108, pp.~171--190, Oct 1957.

\bibitem{Jaynes:SFI90}
E.~T. Jaynes, ``Probability in quantum theory,'' in {\em Complexity, Entropy
  and the Physics of Information} (W.~H. Zurek, ed.), vol.~I, Santa Fe
  Institute, 1990.

\bibitem{Ballentine:RevModPhys70}
L.~E. Ballentine, ``The statistical interpretation of quantum mechanics,'' {\em
  Rev. Mod. Phys.}, vol.~42, pp.~358--381, Oct 1970.

\bibitem{Appleby:FoundPhys05}
D.~M. Appleby, ``Facts, values and quanta,'' {\em Foundations of Physics},
  vol.~35, no.~4, pp.~627--668, 2005.

\bibitem{Carroll:QuantaMag19}
S.~Carroll, ``Where quantum probability comes from,'' {\em Quanta Magazine},
  September 2 2019.

\bibitem{Rovelli:IntJTheoreticalPhys96}
C.~Rovelli, ``Relational quantum mechanics,'' {\em International Journal of
  Theoretical Physics}, vol.~35, pp.~1637--1678, aug 1996.

\bibitem{Fuchs:RevModPhys13}
C.~A. Fuchs and R.~Schack, ``{Quantum-Bayesian coherence},'' {\em Reviews of
  Modern Physics}, vol.~85, pp.~1693--1715, dec 2013.

\bibitem{Frauchiger:ArXiv17}
D.~Frauchiger and R.~Renner, ``A non-probabilistic substitute for the {B}orn
  rule,'' 2017, 1710.05033.

\bibitem{Matsoukas:arXivSM23}
T.~Matsoukas, ``Stochastic processes and statistical mechanics,'' 2023,
  2103.09909.

\bibitem{Zurek:PhysicsToday14}
W.~H. Zurek, ``{Quantum Darwinism, classical reality, and the randomness of
  quantum jumps},'' {\em Physics Today}, vol.~67, pp.~44--50, 10 2014.

\bibitem{Rozanov:77}
Y.~A. Rozanov, {\em Probability Theory: A Concise Course}.
\newblock New Yoork, NY: Dover Pubications, 1977.

\end{thebibliography}
% \bibliographystyle{hieeetr}

%--------------------------------------------------------------------------*
\onecolumngrid
\appendix
\section{Derivations}
%--------------------------------------------------------------------------*
\subsection{Two-state quantum system}
\label{app_2sqs}

The solution to the Schrödinger equation with Hamiltonian 
\begin{gather}
  \mathscr H = 
  \Matrix
    {\alpha+\delta}
    {\beta}
    {\beta}
    {\alpha-\delta}
\end{gather}
and initial state $\ket{\psi_0} = c_1 \ket{1}+c_2\ket{2}$ is
\begin{gather}
\label{app_schrodinger_1}
  \Vector{\psi_1(t)}{\psi_2(t)}
  = 
  \frac{1}{\sqrt{c_1^2+c_2^2}}
  \Matrix{u_{11}}{u_{12}}{u_{21}}{u_{22}}\cdot
  \Vector{c_1}{c_1}
\end{gather}
with 
\begin{gather}
  u_{11} = 
    e^{-\ii \alpha  t/\hbar }
    \left(\cos\theta-\frac{\ii \delta }{\sqrt{\beta^2+\gamma^2}}  
    \sin\theta\right)
  \\
  u_{12} = 
    e^{-\ii \alpha  t/\hbar }
    \left(
      \frac{-\ii \beta}{\sqrt{\beta^2+\gamma^2}}
      \sin\theta
    \right)
  \\
  u_{21} =
    e^{-\ii \alpha  t/\hbar }
     \left(
       -\frac{\ii\beta}{\sqrt{\beta^2+\gamma^2}}
       \sin\theta
     \right)
  \\
\label{app_schrodinger2}
u_{22} = 
    e^{-\ii \alpha  t/\hbar }
    \left(\cos\theta + 
    \frac{\ii \delta}{\sqrt{\beta^2+\gamma^2}}\sin\theta
    \right)
  \\ 
  \theta = \frac{t\sqrt{\beta^2+\delta^2}}{\hbar}.
\end{gather}
By including the normalization factor $\sqrt{c_1^2+c_2^2}$ in Eq.\ (\ref{app_schrodinger_1}) we are free to set $c_1$ and $c_2$ independently of each other. The corresponding probabilities are
\begin{gather}
\label{app:prob_QM12}
  \begin{array}{*2{>{\displaystyle}l}}
      P_1^\text{QM}(\theta) 
      = \psi_1(\theta)\psi_1^*(\theta)
      & = \frac{c_1^2}{c_1^2+c_2^2} \cos^2(\theta ) 
      + \sin ^2(\theta ) 
        \frac{\left(c_1 \beta +c_2 \delta \right)^2}
             {(c_1^2+c_2^2)(\beta^2+\delta^2)}
      \\ 
      P_2^\text{QM}(\theta) 
        = \psi_2(\theta)\psi_2^*(\theta)
        & = \frac{c_2^2}{c_1^2+c_2^2} \cos ^2(\theta )
        + \sin ^2(\theta ) 
          \frac{\left(c_2 \beta -c_1 \delta \right)^2}
               {(c_1^2+c_2^2)(\beta^2+\delta^2)} .
  \end{array}
\end{gather}
Probabilities oscillate with period $T = \pi\hbar/\sqrt{\beta^2+\gamma^2}$ and amplitude that is bounded by their values at $\theta=0$ and $\theta=\pi/2$. For state 1 these bounds are
\begin{gather}
\renewcommand{\arraystretch}{3}
  \begin{array}{>{\displaystyle}ll}
    P^\text{QM}_{1}(0) = \frac{c_1^2}{c_1^2+c_2^2} 
    &\equiv P',
    \\
    P^\text{QM}_{1}(\pi/2) 
    = \frac{(c_1 \delta + c_2\beta)^2}{(c_1^2+c_2^2)(\beta^2+\delta^2)}  
   &\equiv P'' .
  \end{array}
\end{gather}
These are illustrated in Fig.\ \ref{fig1}. 

%--------------------------------------------------------------------------*
\subsection{Classical stochastic process: Connection to Markov}
\label{app_markov}

First we write Eq.\ (\ref{propagation}) in the form
\begin{gather}
\label{app:markov1}
  \begin{array}{ll}
    a_1 = k_{11}a'_1 + k_{12} a'_2;\quad~  & a_{1;0} = c_1\\ 
    a_2 = k_{21}a'_1 + k_{22} a'_2;        & a_{2;0} = c_2
  \end{array} ,
\end{gather}
with $a_i = a_{i;n}$, $a'_i = a_{i;n-1}$. The corresponding probabilities are 
\begin{gather}
   P_i  = \frac{a_i}{a_1+a_2},\quad
   P'_i = \frac{a'_i}{a'_1+a'_2};\quad
   i=1,2 .  
\end{gather}
Combining these results we obtain equations for the evolution of the probabilities:
\begin{gather}
  P_1 = \frac
    {k_{11}P'_1 + k_{12}P'_2}
    {(k_{11} + k_{21})P'_1 + (k_{12} + k_{22})P'_2} ,
    \\
  P_2 = \frac
    {k_{21}P'_1 + k_{22}P'_2} 
    {(k_{11} + k_{21})P'_1 + (k_{12} + k_{22})P'_2} .
\end{gather}
The stationary solution is obtained by setting $P_i=P'_i = P_i^*$  and solving for $P_i^*$. 
Markov behavior is obtained in the special case
\begin{gather}
   k_{11}+k_{21}=k_{12}+k_{22} = 1.
\end{gather}
The probabilities then reduce to
\begin{gather}
\label{app:markov2}
  \begin{array}{*2{>{\displaystyle}l}}
    P_1 = k_{11}P'_1 + k_{12}P'_2 \\ 
    P_2 = k_{21}P'_1 + k_{22}P'_2
  \end{array}
  \Rightarrow
    \Vector{P_1}{P_2}_n
    = 
    \mathscr K
    \Vector{P_1}{P_2}_{n-1}
\end{gather}
whose solution is
\begin{gather}
\label{app:markov3}
    \Vector{P_1}{P_2}_n
    =
    \mathscr K^n
    \Vector{P_1}{P_2}_0 
\end{gather}
In the Markov case the $k_{ij}$ are properly normalized transition probabilities.

%--------------------------------------------------------------------------*
\subsection{Quantum stochastic process: Conditions for sustained oscillations}
\label{app_oscillations}

We write Eq.\ (\ref{propagation}) more explicitly as
\begin{gather}
  \begin{array}{ll}
    a_{1;n} = k_{11}a_{1;n-1} + k_{12} a_{2;n-1};\quad~  & a_{1;0} = c_1\\ 
    a_{2;n} = k_{21}a_{1;n-1} + k_{22} a_{2;n-1};        & a_{2;0} = c_2
  \end{array},
\end{gather}
and calculate the probabilities of state 1 at the initial state ($P_{1;0}$) and after the second transition ($P_{1;2}$):
\begin{gather}
\label{app:prob_QS}
  \begin{array}{>{\displaystyle}l}
    P^\text{QSP}_{1;0} = \frac{c_1^2}{c_1^2+c_2^2}
    \\ 
    P^\text{QSP}_{1;2} = \frac
    {
      \left(k_{11} \left(c_1 k_{11}+c_2 k_{12}\right)+k_{12} \left(c_1 k_{21}+c_2 k_{22}\right)\right){}^2
    }
    {
      \left(k_{11} \left(c_1 k_{11}+c_2 k_{12}\right)+k_{12} \left(c_1 k_{21}+c_2 k_{22}\right)\right){}^2+\left(k_{21} \left(c_1 k_{11}+c_2 k_{12}\right)+k_{22} \left(c_1 k_{21}+c_2 k_{22}\right)\right){}^2
    } .
  \end{array}
\end{gather}
A necessary condition for oscillations is $P^\text{QSP}_{1;0} = P^\text{QSP}_{1;2}$ for all $c_1$, $c_2$. This must also be true for $c_1=1$, $c_2=0$, in which case the condition $P_{1;0} = P_{1;2}$ is equivalent to (Mathematica 13.2.0.0)
\begin{gather}
\label{app_cond1}
  \left(k_{11}=-k_{22}\land k_{22}^2+k_{12} k_{21}\neq 0\right)\lor \left(k_{21}=0\land k_{11}\neq 0\right)  . 
\end{gather}
%
% (The result was obtained using the \textsf{Reduce} function in \textit{Mathematica} 13.0.1.0.)
The case $k_{21}=0$ is rejected because it gives $P^\text{QSP}_1=1$ at all times. 
With $k_{11}=-k_{22}$ in Eq.\ (\ref{app:prob_QS}) we confirm that $P^\text{QSP}_{1;0} = P^\text{QSP}_{1;3}$ is indeed satisfied for all $c_i$. 

%--------------------------------------------------------------------------*
\subsection{Relationship between transition matrix and the Hamiltonian}
\label{app_equivalence}

To obtain equivalence between the quantum model and the quantum stochastic process we  match their probabilities at time zero and time equal to one-half period:
\begin{gather}
   P^\text{QM}_{1}(0)     = P^\text{QSP}_{1;0},\quad
   P^\text{QM}_{1}(\pi/2) = P^\text{QSP}_{1;1} . 
\end{gather}
The first equation is always satisfied. The second equation fixes the relationship between the elements of the Hamiltonian and the transition rates of the QSP. To obtain this relationship we set $c_2=1-c_1$,  write these probabilities in the form
\begin{gather}
  P^\text{QM}_{1;1} = \frac{A_2 c_1^2 + A_1 c_1 + A_0}{{B_2 c_1^2 + B_1 c_1 + B_0}},\quad
  P^\text{QSP}(\pi/2)  = \frac{D_2 c_1^2 + D_1 c_1 + D_0}{{E_2 c_1^2 + E_1 c_1 + E_0}},\quad
\end{gather}
and require the coefficients of equal powers of $c_1$ in the numerators and  denominators to be proportional to each other:
\begin{gather}
  (A_2,A_1,A_0,B_2,B_1,B_0) = \frac{1}{\lambda^2} (C_2,C_1,C_0,D_2,D_1,D_0) ,
\end{gather}
where $\lambda\neq 0$ is a factor independent of $c_1$. This leads to the conditions (Mathematica 13.2.0.0)
\begin{gather}
\label{app_cond2}
  k_{11} =-k_{22} = \pm \lambda\delta,\quad
  k_{12} = k_{21} = \pm \lambda \beta . 
\end{gather}
with the signs chosen in tandem. Since $\lambda$ may be chosen positive or negative the $\pm$ signs may be dropped. 

The quantum stochastic process produces periodic oscillations under the weaker condition (\ref{app_cond1}). Equivalence with quantum mechanics requires as an additional condition Eq.\ (\ref{app_cond2}), which makes the transition matrix Hermitian.  

%--------------------------------------------------------------------------*
\subsection{Normalization and unitary propagation}
\label{app_unitary}

We combine Eqs.\ (\ref{app_cond1}) and (\ref{app_cond2}) to express the conditions on $k_{ij}$ in the form
\begin{gather}
   k_{11} = - k_{22} = K,\quad
   k_{12} = k_{21} = L  
\end{gather}
and write the transition matrix as 
\begin{gather}
   \mathscr K = \Matrix{K}{\hphantom{-}L}{L}{-K} . 
\end{gather}
We may show recursively that $\mathscr K^n$ is given by 
\begin{gather}
  \mathscr K^n = 
    \begin{cases}
      \Matrix
        {(K^2+L^2)^{n/2}}
        {0}
        {0}
        {-(K^2+L^2)^{n/2}} 
      & \text{if $n$ is even} \\\\
      \Matrix
        {K(K^2+L^2)^{(n-1)/2}}
        {L(K^2+L^2)^{(n-1)/2}}
        {L(K^2+L^2)^{(n-1)/2}}
        {-K(K^2+L^2)^{(n-1)/2}} 
      & \text{if $n$ is odd} .
   \end{cases}
\end{gather}
Using this result along with Eq.\ (\ref{propagator}) we  obtain the normalization factor of the probability,
\begin{gather}
  A^2_n \equiv 
  a_{1;n}^2 + a_{2;n}^2 = (c_1^2+c_2^2)(K^2+L^2)^n  , 
\end{gather}
which is true for all $n$, even or odd. 
The probability of state $i$ at $t_n = n \tau$ is
\begin{gather}
   p_{i;n} = \frac{a_{1;n}^2}{A^2_n} = \frac{a_{1;n}^2}{(c_1^2+c_2^2)(K^2+L^2)^n}   
\end{gather}
and with the normalizations
\begin{gather}
   K^2+L^2 = c_1^2+c_2^2 = 1,
\end{gather}
this becomes $p_{i;n} = a_{i;n}^2$ and satisfies $p_{1;n}+p_{2;n}=1$ for all $n$. 
Under this normalization the evolution matrix simplifies to
\begin{gather}
  \mathscr K^n = 
    \begin{cases}
      \Matrix
        {1}
        {0}
        {0}
        {1} 
      & \text{if $n$ is even} \\\\
      \Matrix
        {K}
        {L}
        {L}
        {-K} 
      & \text{if $n$ is odd} .
   \end{cases}
\end{gather}
%
%--------------------------------------------------------------------------*
\subsection{Wave function and the Schrödinger equation}
\label{app_wavefunction}

The QSP gives the probability of observation at discrete times $t_n = n \tau$, where $\tau$ is the mean time between transitions. We will construct a smooth interpolation of the discrete probability as a continuous function of time. We define the complex vector
\begin{gather}
\label{app:psi_def}
  \Vector{\psi_1(t)}{\psi_2(t)}
  =
  \left\{
    X(t)\Matrix
        {1}
        {0}
        {0}
        {1}
     - \ii 
     X(t+\tau)\Matrix
        {K}
        {L}
        {L}
        {-K}
  \right\} 
  \Vector{c_1}{c_2} , 
\end{gather}
where $X(t)$ is a periodic function between 1 and 0 with period $2\tau$:
\begin{gather}
\label{app:X_def}
  X\left(t = n \tau\right) = 
  \begin{cases}
    1 & \text{if $n$ is even},
    \\ 
    0 & \text{if $n$ is odd} . 
  \end{cases}
\end{gather}
As we can confirm, the square modulus of $\psi_i$ at $t=n\tau$ gives the probability $P_i$ for all $n$:
\begin{gather}
\label{app:condition12}
  \psi_1(t) \psi_1^*(t) = P_{1}(t) , \quad
  \psi_2(t) \psi_2^*(t) = P_{2}(t) . 
\end{gather}
To use this equation to calculate the probability for all $t$ we must determine $X(t)$ such that the normalization condition 
\begin{gather}
\label{app:condition3}
  \psi_1(t) \psi_1^*(t) + \psi_2(t) \psi_2^*(t) = 1  
\end{gather}
is satisfied for all $t$ and all $c_i$. We set $c_2=1-c_1$ and substitute into Eq.\ (\ref{app:condition3}) to obtain
\begin{gather}
  \Big(X^2(t) + X^2(t+\tau/2) - 1\Big)
  \Big(c_1^2+(1-c_1)^2\Big) 
  = 0 . 
\end{gather}
For this to be true for all $c_1$ we must have
\begin{gather}
\label{app:X_condition}
  X^2(t) + X^2(t+\tau/2) = 1 .
\end{gather}
From Eqs.\ (\ref{app:X_def}) and (\ref{app:X_condition}) we conclude
\begin{gather}
\label{app:X}
  X(t) = \cos\left(\frac{\pi t}{\tau}\right) . 
\end{gather}
Returning this result into Eq.\ (\ref{app:psi_def}) we obtain
\begin{gather}
  \Vector{\psi_1}{\psi_2}
  = 
  \Matrix
  {\cos\theta - \ii L \sin\theta}
  {-\ii L\sin\theta}
  {-\ii L\sin\theta}
  {\cos\theta + \ii L \sin\theta}
  \Vector{c_1}{c_2} ,  
\end{gather}
with $\theta=\pi t/\tau$. 
This is the same as the solution of the Schrödinger equation under the substitutions
\begin{gather}
   K =  \frac{\beta} {\sqrt{\beta^2+\delta^2}},\quad
   L = -\frac{\delta}{\sqrt{\beta^2+\delta^2}},\quad
   \tau = \frac{\hbar}{\sqrt{\beta^2+\delta^2}} . 
\end{gather}
%
%--------------------------------------------------------------------------*
\subsection{The fair coin}
\label{app_coin}
Setting $\delta=0$, $c_1=1$, $c_2=0$ in Eqs.\ (\ref{app:prob_QM12}) we obtain
\begin{gather}
   P_1(t) = \cos^2(\theta), \quad
   P_2(t) = \sin^2(\theta) .
\end{gather}
This is the same as Eq.\ (\ref{fair_coin}) with $\phi/2=\theta =t\beta/\hbar$. The corresponding Hamiltonian in this case is
\begin{gather}
   \mathscr H = \Matrix{0}{\beta}{\beta}{0}   
\end{gather}
with $\beta$ either positive or negative. The transition matrix of the classical stochastic process that represents the fair coin is
\begin{gather}
   \mathscr K = \Matrix{0}{L}{L}{0}
\end{gather}
with $L$ arbitrary but positive. We then have
\begin{gather}
  \mathscr K^n = 
  \begin{cases}
    \Matrix{L^n}{0}{0}{L^n} & \text{if $n$ even} \\ \\
    \Matrix{0}{L^n}{L^n}{0} & \text{if $n$ odd}
  \end{cases}
\end{gather}
With $c_1=1$, $c_2=0$, this produces alternating probabilities between 1 and 0 for all $L$ and produces the results in Fig.\ \ref{fig3}d. 
%
%--------------------------------------------------------------------------*
\end{document}